\documentclass[manuscript,tighten,times]{aastex63}

\newcommand{\msol}{\mbox{$M_\odot$}}

\newcommand {\Lo}{{L}_\odot}
\newcommand{\HI}{H\,{\sc i} }
\newcommand{\hi}{\ifmmode{\rm HI}\else{H\/{\sc i}}\fi} 
\newcommand {\kms}{\ifmmode{\rm km \, s^{-1}}\else{$\rm km \, s^{-1}$}\fi} 
\newcommand{\Msun} {M_{\sun}}

\begin{document}
\title{Discovery of an isolated dark dwarf galaxy in the nearby universe}

\correspondingauthor{Jin-Long Xu and Ming Zhu}
\email{xujl@bao.ac.cn,  mz@nao.cas.cn}

\author{Jin-Long Xu}
\affiliation{National Astronomical Observatories, Chinese Academy of Sciences, Beijing 100101, People's Republic of China}
\affil{Guizhou Radio Astronomical Observatory, Guizhou University, Guiyang 550000, People's Republic of China}
\affil{CAS Key Laboratory of FAST, National Astronomical Observatories, Chinese Academy of Sciences, Beijing 100101, People's Republic of China}

\author{Ming Zhu}
\affiliation{National Astronomical Observatories, Chinese Academy of Sciences, Beijing 100101, People's Republic of China}
\affil{Guizhou Radio Astronomical Observatory, Guizhou University, Guiyang 550000, People's Republic of China}
\affil{CAS Key Laboratory of FAST, National Astronomical Observatories, Chinese Academy of Sciences, Beijing 100101, People's Republic of China}

\author{Naiping Yu}
\affiliation{National Astronomical Observatories, Chinese Academy of Sciences, Beijing 100101, People's Republic of China}
\affil{Guizhou Radio Astronomical Observatory, Guizhou University, Guiyang 550000, People's Republic of China}
\affil{CAS Key Laboratory of FAST, National Astronomical Observatories, Chinese Academy of Sciences, Beijing 100101, People's Republic of China}

\author{Chuan-Peng Zhang}
\affiliation{National Astronomical Observatories, Chinese Academy of Sciences, Beijing 100101, People's Republic of China}
\affil{Guizhou Radio Astronomical Observatory, Guizhou University, Guiyang 550000, People's Republic of China}
\affil{CAS Key Laboratory of FAST, National Astronomical Observatories, Chinese Academy of Sciences, Beijing 100101, People's Republic of China}

\author{Xiao-Lan Liu}
\affiliation{National Astronomical Observatories, Chinese Academy of Sciences, Beijing 100101, People's Republic of China}
\affil{Guizhou Radio Astronomical Observatory, Guizhou University, Guiyang 550000, People's Republic of China}
\affil{CAS Key Laboratory of FAST, National Astronomical Observatories, Chinese Academy of Sciences, Beijing 100101, People's Republic of China}

\author{Mei Ai}
\affiliation{National Astronomical Observatories, Chinese Academy of Sciences, Beijing 100101, People's Republic of China}
\affil{Guizhou Radio Astronomical Observatory, Guizhou University, Guiyang 550000, People's Republic of China}
\affil{CAS Key Laboratory of FAST, National Astronomical Observatories, Chinese Academy of Sciences, Beijing 100101, People's Republic of China}

\author{Peng Jiang}
\affiliation{National Astronomical Observatories, Chinese Academy of Sciences, Beijing 100101, People's Republic of China}
\affil{Guizhou Radio Astronomical Observatory, Guizhou University, Guiyang 550000, People's Republic of China}
\affil{CAS Key Laboratory of FAST, National Astronomical Observatories, Chinese Academy of Sciences, Beijing 100101, People's Republic of China}

\begin{abstract}
Based on a new \HI survey using the Five-hundred-meter Aperture Spherical radio Telescope (FAST), combined with the Pan-STARRS1 images, we identified an isolated \HI cloud without any optical counterpart, named FAST J0139+4328. The newly discovered \HI cloud appears to be a typical disk galaxy since it has a double-peak shape in the global \HI profile and  an S-like rotation structure in the velocity-position diagram.   Moreover, this disk galaxy has an extremely low absolute  magnitude ($M_{\rm B}>$-10.0 mag) and  stellar mass ($<$6.9$\times10^{5} M_{\odot}$). Furthermore, we obtained that the \HI mass of this galaxy is (8.3$\pm$1.7)$\times10^{7} \Msun$, and  the dynamical mass to total baryonic mass ratio is 47$\pm$27, implying that dark matter dominates over baryons in FAST J0139+4328. These findings provide observational evidence that FAST J0139+4328 is an isolated dark dwarf galaxy with a redshift of $z=0.0083$. This is the first time that an isolated dark galaxy has been detected in the nearby universe.
\end{abstract}


\section{Introduction} \label{sec:intro}
One of the puzzles in extragalactic astronomy is the disparity between the $\Lambda$ cold dark matter ($\Lambda$CDM)  predictions and observations of dwarf galaxy numbers \citep{Kauffmann1993,Tollerud2008}. It is also often referred to as the ``missing satellite problem'' \citep{Moore1999}. With the detection of a large number of dwarf galaxies and  the improvement of theoretical models, the disparity has significantly decreased, and even this is no longer a missing satellite problem \citep{Sales2022}. However, the number of ultra-faint galaxies ($M_{\ast}<10^{5}$ $M_{\odot}$) remains largely unconstrained \citep{Bullock2017,Sales2022}. For more massive galaxies, the various numerical simulations indicated a tight relation between halo mass and stellar mass with little scatter \citep{Sawala2016,Behroozi2013}. This relation did not hold true for ultra-faint galaxies, because the scatter was predicted to be greater. It indicates that a significant amount of theoretical uncertainty still exists. On the other side, a significant number of ultra-faint galaxies  have not been detected in observations. Therefore, determining whether theoretical predictions and observations are consistent remains difficult.  Dark galaxies are thought to be dark matter  haloes whose gas failed to form stars \citep{Roman2021}. The existence of dark galaxies could offer a standard way to explain the problem for the absence of ultra-faint galaxies. Furthermore, the discovery of dark galaxies is crucial to understanding galaxy formation since gas-rich dark galaxies may reflect the earliest stage of galaxy formation.

The search for dark galaxies has been going on for many years, but none of the candidates appear to be ideal \citep{Taylor2016}. The southwestern (SW) clump of  HI 1225+01 and  the VIRGOHI 21  in the Virgo cluster are the best
dark-galaxy candidates. However, both these candidates are just located at the tidal tails, for which they could be considered as possible tidal debris \citep{Turner1997,Duc2008}. Because it is difficult to determine whether the detected candidates are dark galaxies or just unusual tidal debris, the ideal candidates for  a dark galaxy should be located far  away from other massive galaxies,  groups, and clusters.  Blind  neutral hydrogen (\HI) surveys are one of the most efficient methods for searching the dark galaxies.  However, both the \HI Parkes All Sky Survey (HIPASS) and Arecibo Legacy Fast ALFA (ALFALFA)  surveys failed to find any isolated dark galaxies \citep{Doyle2005,Haynes2011}. Even if some previously discovered candidates were later found to be only a few almost dark galaxies  based on deeper optical observations \citep{Ball2018,Leisman2021}. \citet{Davies2006} predicted that dark galaxies exist, but a significant portion of these galaxies may have masses less than 10$^{7} \Msun$ and velocity widths under 40 \kms. Thus, both high sensitivity and high velocity resolution play a key role in searching for dark galaxies in a new \HI survey. 

Using the Five-hundred-meter Aperture Spherical radio Telescope \citep{Jiang2019,Jiang2020}, we set out to carry out a FAST extragalactic \HI survey, which is a time-filler project when there are no other programs in the FAST observing queue. A major scientific objective for this survey is to search for dark and weak galaxies taking advantage of high sensitivity of FAST and with a relatively high velocity resolution.  In this Letter, based on the \HI survey using FAST, we have discovered an isolated dark dwarf galaxy in the nearby universe.

\section{Observation and data processing}
For a new \HI survey using FAST, the 19-beam array receiver system in dual polarization mode is used as  the front end. It formally works in the frequency range from 1050 MHz to 1450 MHz. For the backend, we choose the Spec(W) spectrometer that has 65,536 channels covering a bandwidth of 500 MHz for each polarization and beam, resulting in a frequency resolution of 7.629 kHz and corresponding to a  velocity resolution of 1.6 \kms at z=0. The FAST \HI survey uses the drift scan mode. An interval between two adjacent parallel scans  in decl. is $\sim$1.14$^{\prime}$. Besides, we set an integration time of 1 second per spectrum. The system temperature was around 25 K. For intensity calibration to antenna temperature ($T_{\rm A}$),  a noises signal with amplitude of 10 K was injected for 1 s every 32 s. The half-power beam width (HPBW) is $\sim$2.9$^{\prime}$ at 1.4 GHz for each beam. The pointing accuracy of the telescope was better than 10$^{\prime\prime}$. The FAST \HI observations of FAST J0139+4328 in the new survey were carried out on  August 2, 2021 (region ID: DEC+433059-6). The detailed data reduction is similar to \citet{Xu2021}.  According to \citet{Zhang2022}, we flag the radio frequency interferences (RFIs). The RFI contamination rate is minimal in the FAST data used in this paper. The beamwidth after gridding is $\sim$3.1$^{\prime}$.  A gain $T_{\rm A}/\it S_{v}$ has been measured to be about 16 K Jy$^{-1}$. The  measured relevant main beam gain $T_{\rm B}/\it S_{v}$ is about 21 K Jy$^{-1}$ at 1.4 GHz for each beam, where $T_{\rm B}$ is the brightness temperature.  To construct a highly-sensitive  \HI image, the velocity resolution of the FAST data is smoothed to 4.8 $\kms$. The mean noise RMS in the observed image is $\sim$0.9 mJy beam$^{-1}$.

\section{Results}
\label{sect:results}
So far, we have detected some new \HI clouds in local universe based on the new FAST extragalactic \HI survey data.  Among these detected clouds, we found an isolated cloud,  named FAST J0139+4328 according to its coordinates. The isolated clouds are typically thought to have no relatively massive galaxies within a radius of 100 kpc \citep{Taylor2017}.
Under the assumption that the gas in galaxies is optically thin for the \HI line, the column density $N(x,y)$ in each pixel can be estimated as $N(x,y)= 1.82\times10^{18}\int T_{\rm B}dv$,  where $dv$ is the velocity width in \kms. 
Figure~\textcolor{blue}{1a}  shows the \HI column-density map of FAST J0139+4328,  overlaid on the Pan-STARRS1 g-band image in colorscale.  From the \HI emission, we can see that FAST J0139+4328 has a compact structure. In Fig. \textcolor{blue}{1b}, we display the global \HI profile of FAST J0139+4328. The \HI profile seems to exhibit a double-peak shape, which is generally thought to be caused by the flat rotation curve of a disk galaxy. We performed a BusyFit fitting to the profile to determine the  system velocity ($V_{\rm sys}$), total flux ($S_{v}$), line widths at 50\% of the peak flux ($W_{50}$) and at 20\% of the peak flux ($W_{20}$). The busy function is a new analytic function for describing the integrated \HI spectral profile of galaxies \citep{Westmeier2014}. The parameters obtained by the fitting are listed  in Table \ref{tab:prop}.

\begin{figure*}
\centering
\includegraphics[width=0.32\textwidth]{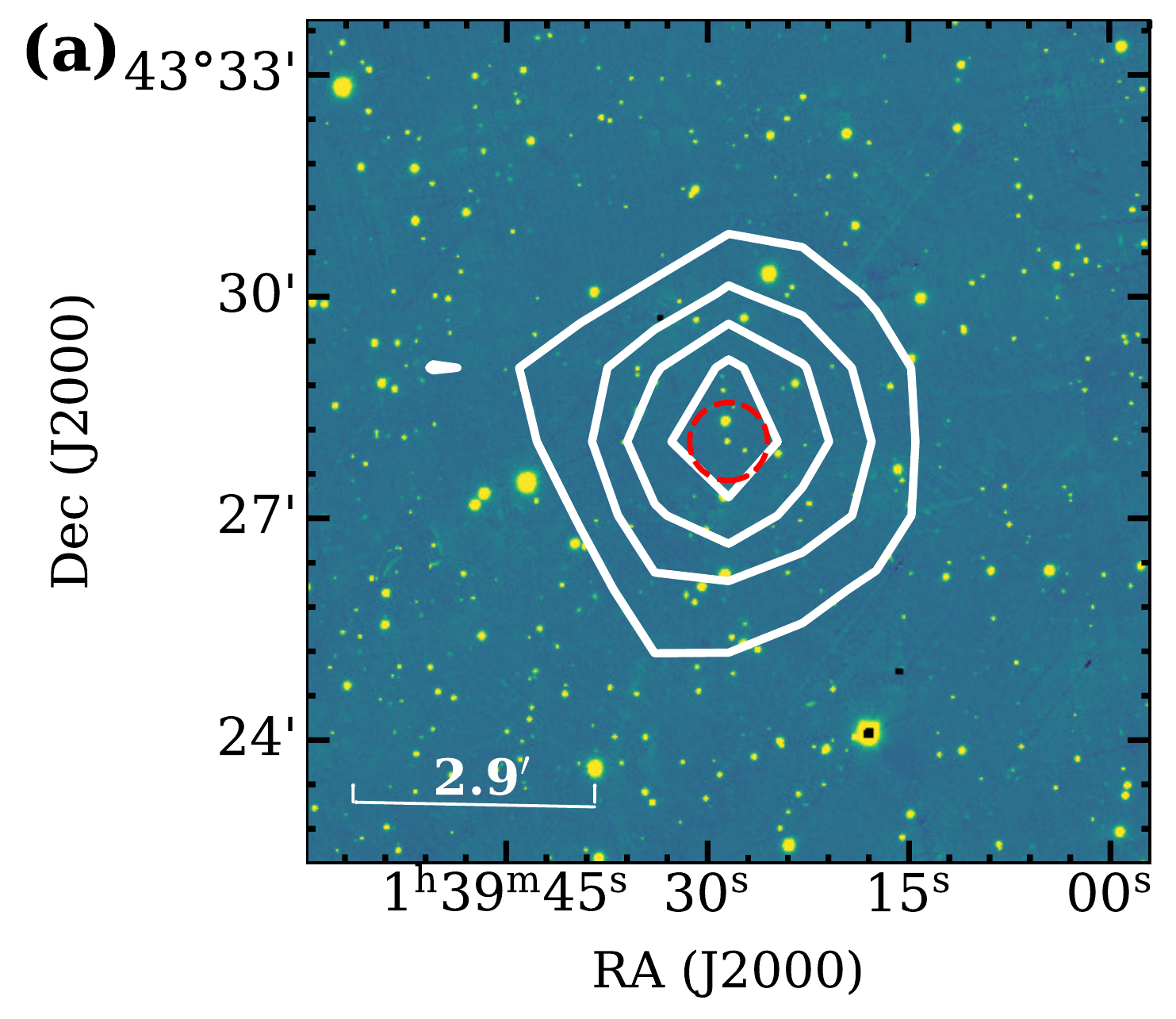}
\includegraphics[width=0.32\textwidth]{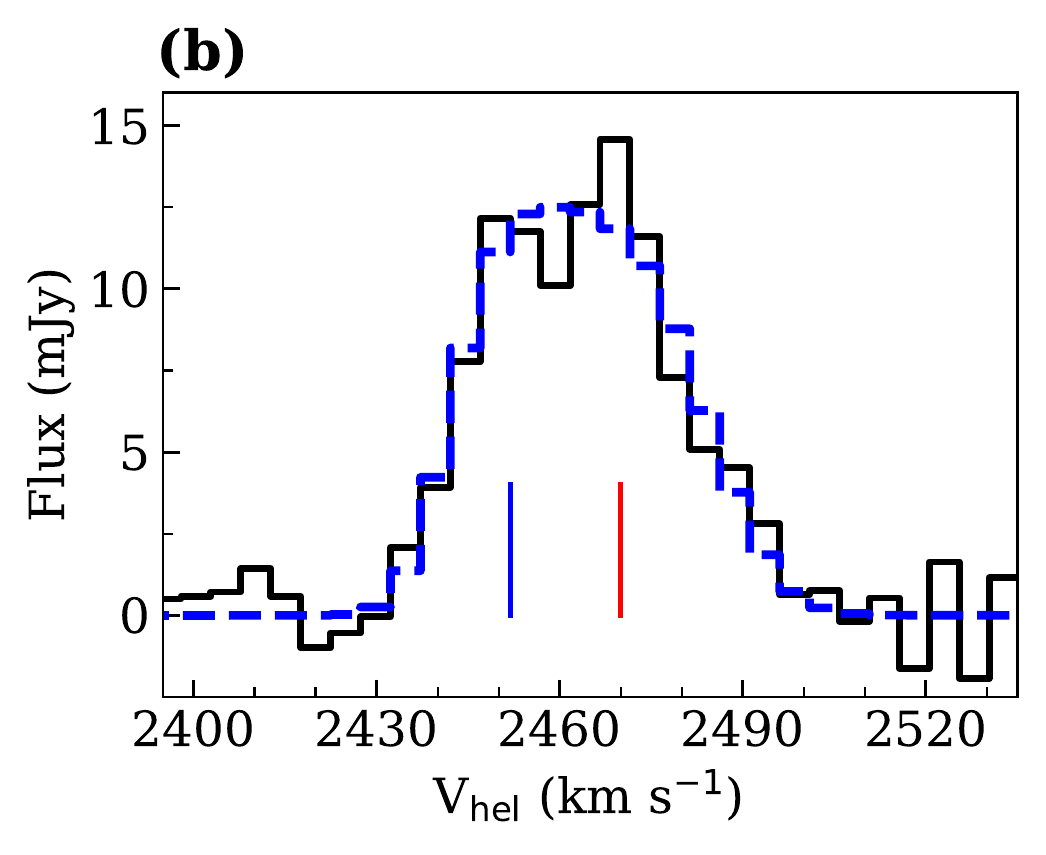}
\includegraphics[width=0.31\textwidth]{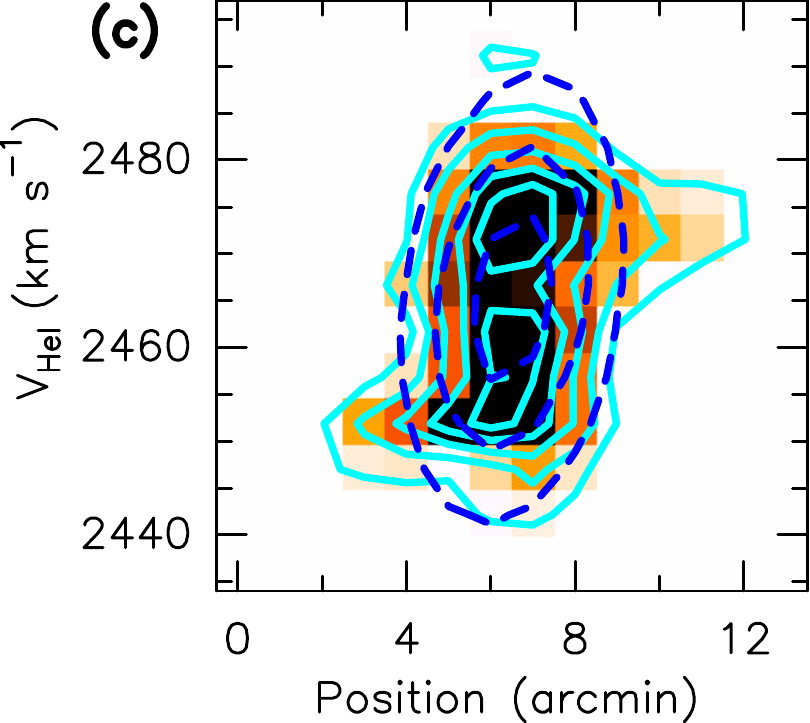}
\caption{Morphology of dark galaxy FAST J0139+4328. {\bf a}, \HI column-density map from the FAST observation shown in  white contours overlaid on the Pan-STARRS1 g-band image in colorscale. The white contours begin at 3.0$\times$10$^{18}$ cm$^{-2}$ (3$\sigma$) in steps of 1.9$\times$10$^{18}$ cm$^{-2}$. The red dashed circle represents the three times the expected $D_{25}$. {\bf b}, global \HI profile of FAST J0139+4328. The blue dashed line indicates the BusyFit fitting result. The red and blue vertical lines mark the positions of double peak.  {\bf c}, position-velocity (PV) diagram in colorscale overlaid with the cyan contours. The cyan contours begin at 2$\sigma$ (1.8 mJy beam$^{-1}$) in steps of 1$\sigma$. The blue dashed contours represent the PV diagram from the TiRiFiC fitting.}
\label{fig:Dark_galaxy-HI}
\end{figure*}

Figure \textcolor{blue}{1c}  shows the position-velocity (PV) diagram in cyan contours for FAST J0139+4328. The extracted PV diagram passes through the center of FAST J0139+4328 with a length of 14.0$^{\prime}$ along  a position angle of 151$^{\circ}$. From the PV diagram, we see that FAST J0139+4328 seems to have two velocity components, which are associated with the double-peak shape in its global \HI profile. In addition, the whole PV diagram displays an S-like structure, which is a typical characteristics of disk galaxies. Based on the presence of the global \HI profile, and  S-like rotation structure,  we suggest that the \HI cloud FAST J0139+4328 is a newly discovered disk galaxy.  Gas in galaxies mainly consists of \HI and helium.  The \HI gas mass of FAST J0139+4328 can be estimated as  $M_\mathrm{\hi}=2.36\times10^{5}D^{2}{\int}S_{v}dv$,  where  ${\int}S_{v}dv$ is the integrated {\HI} flux in Jy km s$^{-1}$, and $D$ is the adopted distance in Mpc to FAST J0139+4328. We derived that the total $S_{v}$ is 424$\pm$20 mJy km s$^{-1}$ from the global \HI profile of FAST J0139+4328, and the system velocity $V_{\rm sys}$ is 2464.4$\pm$0.8 km s$^{-1}$, corresponding to redshift $z=0.0083$. Using the Cosmicflows-3 method \citep{{Kourkchi2020}},  the distance to FAST J0139+4328 can be estimated to be 28.8$\pm$2.9 Mpc.  Finally, for FAST J0139+4328, we obtained that $M_{\hi}$  is (8.3$\pm$1.7)$\times10^{7} \Msun$.

\begin{figure*}
\centering
\includegraphics[width=.7\textwidth]{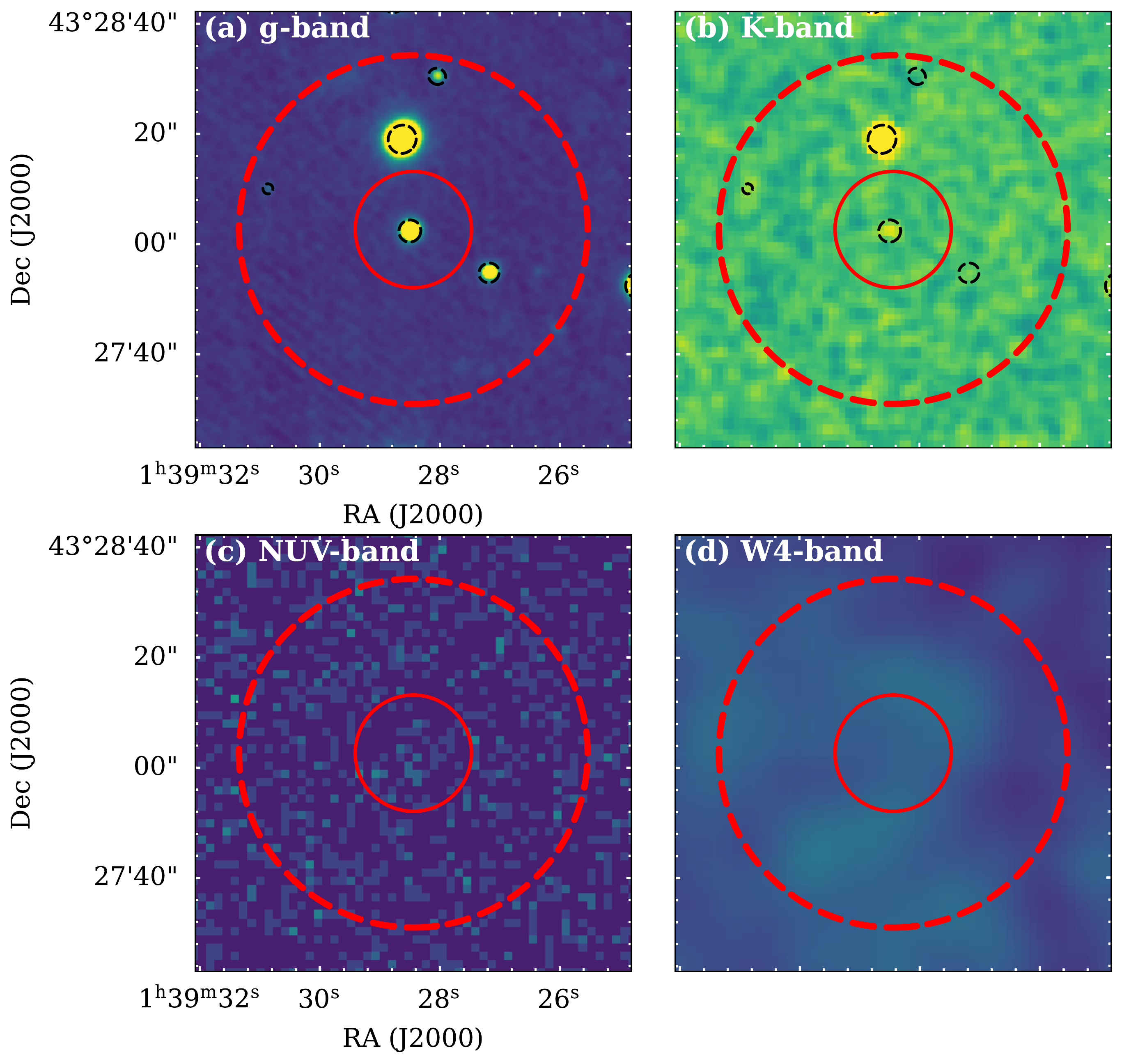} 
\caption{Images of dark galaxy FAST J0139+4328 in different bands. {\bf a}, optical emission from the Pan-STARRS1 g-band data. {\bf b,} near infrared K-band emission from the 2MASS data. {\bf c},  near ultraviolet (NUV) emission from the GALEX data. {\bf d}, mid-infrared emission (W4, 22 $\mu m$) from the WISE data. Fluxes in each band are on an arbitrary unit. The red solid and dashed circles represent the expected $D_{25}$ and the three times $D_{25}$ in each panel, respectively. The black ellipses mark optical stars in the Galaxy identified from the Gaia star catalog.}
\label{fig:Dark_galaxy_bands}
\end{figure*}

The Pan-STARRS1 survey is a 3$\pi$ steradian survey with a medium depth in 5 bands \citep{Chambers2016}. To investigate the presence of stellar components  that  could be associated with galaxy FAST J0139+4328, we use the Pan-STARRS1 r-band and g-band archive images.  Figure~\textcolor{blue}{1a}  shows the Pan-STARRS1 g-band mosaic image in colorscale. Interestingly, we have not detected any extended optical emission in the \HI plane of this galaxy. But we can make some restrictions on the results of the optical observation. Previous work on galaxies discovered a tight relation between the \HI mass ($M_{\rm HI}$) and the optical disc diameter ($D_{25}$): log$_{10}(M_{\rm HI})=a+b\times$ log$_{10}(D_{25})$,  where $D_{25}$ are measured at 25 mag arcsec$^{-2}$ level in B-band \citep{Solanes1996,Broeils1997,Toribio2011}. The coefficients for normal spiral galaxies can be adopted as $a=7.0$ and $b=1.95$ \citep{Broeils1997}. Using the obtained $M_{\rm HI}$,  we expect the  $D_{25}$ value to be $\sim$2.9 kpc (21$^{\prime\prime}$) for FAST J0139+4328.  Figure~\textcolor{blue}{2a} shows zoom-in image in the Pan-STARRS1 g-band, which is three times the expected $D_{25}$ size, as indicated by the red dashed circle. The Gaia data can provide the most precise parallaxes for the identified stars. Through the parallax, we can obtain the relatively accurate distance of these stars. Using the Gaia source catalog \citep{Gaia2016}, we identified five bright stars, which are marked in the black dashed circles. Among these stars, the farthest one with a parallax distance of $\sim$13 kpc  is located within the predicted $D_{25}$ region.   As a result, we were able to confirm that the bright stars in the zoom-in image are some Galaxy stars. Yet we have not detected an optical counterpart for galaxy FAST J0139+4328.  The 2MASS K-band  and GALEX NUV-band images are  often used to determine whether there has been any new star formation recently \citep{Skrutskie2006,Martin2005}. We also did not detect  any sign of star formation in galaxy FAST J0139+4328 based on the images in Figs.~\textcolor{blue}{2b} and ~\textcolor{blue}{2c}. 

Furthermore, strong $L$-band emission without an optical counterpart is detected as those of radio recombination lines (RRL) from a Galaxy H II region \citep{Zhang2021} or  OH megamasers from a luminous/ultra-luminous infrared galaxy \citep{Zhang2014,Hess2021,Glowacki2022}. However, using the Simbad and NED database, we are unable to locate an infrared galaxy in the three times the expected $D_{25}$ region.  We also did not identify a luminous infrared galaxy in the WISE W4 image, as shown in Fig.~\textcolor{blue}{2d}. Hence, we exclude the possibility that FAST J0139+4328 is  an OH megamaser emission from a luminous infrared galaxy. In addition, if the identified FAST J0139+4328 is an H II region, we should be able to detect the radio continuous emission and UV emission. From the NRAO VLA Sky Survey (NVSS) archived images and UV image in  Fig.~\textcolor{blue}{2c}, we did not detect the corresponding emission. Moreover, the central frequency of this detection line in FAST J0139+4328 is about 1419.626 MHz. H166$\alpha$ is the radio recombination line closest to this signal, and its rest frequency is 1424.734 MHz. The frequency difference between FAST J0139+4328 and H166$\alpha$ is 5.108 MHz, corresponding to a velocity difference of about 1070 \kms, indicating that this detection line  does not correspond to a Galactic redshift H II region. Hence we rule out the possibility that FAST J0139+4328 is some RRL emission from an H II region. Since no counterpart of RRL or OH megamaser emission is identified in this region,  we suggest that FAST J0139+4328 is likely to be an isolated dark galaxy.

\begin{table}
\centering
\caption{\small Measured and derived  properties of dark galaxy FAST J0139+4328. We list: Equatorial coordinates (R.A., Decl.); line widths  at 50\% of the peak flux ($W_{50}$); line widths  at 20\% of the peak flux ($W_{20}$);  system velocity ($V_{\rm sys}$); distance ($D$); effective radius ($R_{\rm eff}$); inclination angle ($i$); position angle ($PA$); rotation velocity ($V_{\rm rot}$); velocity dispersion ($\sigma_{v}$); \HI gas mass ($M_{\rm \hi}$);  dynamic mass ($M_{\rm dyn}$);  reddening ($E(B-V)$); $g$-band, $r$-band and $B$-band apparent magnitudes ($m_{\rm g}$, $m_{\rm r}$, $m_{\rm B}$), which are corrected for Galactic extinction; absolute $g$-band, $r$-band and $B$-band magnitudes (M$_{\rm g}$, M$_{\rm r}$, M$_{\rm B}$); $B$-band luminosity ($L_{B}$); stellar mass ($M_{\rm \star}$).}
\label{tab:prop}
\setlength{\tabcolsep}{27pt}
\begin{tabular}{lcccc}
\noalign{\vspace{5pt}}\hline\hline\noalign{\vspace{5pt}}
 Name & FAST J0139+4328  \\
\noalign{\vspace{5pt}}\hline\hline\noalign{\vspace{5pt}}
R.A. & 01$^{\rm h}$39$^{\rm m}$28.4$^{\rm s}$  \\
Decl. & 43$^{\rm \circ}$28$^{\rm \prime}$02.6$^{\rm \prime\prime}$ \\
$W_{\rm 50}$ ($\kms$) & 38.9$\pm$1.9  \\
$W_{\rm 20}$ ($\kms$) & 51.1$\pm$2.0   \\
$V_{\rm sys}$ ($\kms$) & 2464.4$\pm$0.8 \\
$D$ (Mpc) & 28.8$\pm$2.9 \\
$R_{\rm eff}$ (kpc) & 23.9$\pm$2.4  \\
$i$ (deg) & 27.0$\pm$11.6  \\
$PA$ (deg) & 151.0$\pm$3.0  \\
$V_{\rm rot}$ ($\kms$) & 26.9$\pm$9.4  \\
$\sigma_{v}$ ($\kms$) & 7.8$\pm$1.4  \\
$M_{\rm \hi}$ (\msol) & (8.3$\pm$1.7)$\times10^{7}$  \\
$M_{\rm dyn}$ (\msol) & (5.1$\pm$2.8)$\times10^{9}$  \\
$E(B-V)$       & 0.08  \\
$m_{\rm g}$ (mag) & $>$22.0  \\
$m_{\rm r}$ (mag) & $>$21.7  \\
$m_{\rm B}$ (mag) & $>$22.4  \\
$M_{\rm g}$ (mag) & $>$-10.3 \\
$M_{\rm r}$ (mag) & $>$-10.7 \\
$M_{\rm B}$ (mag) & $>$-10.0 \\
$L_{\rm B}$ ($\Lo$) & $<$1.4$\times10^{6}$ \\
$M_{\rm \star}$ (\msol) & $<$6.9$\times10^{5}$ \\
\noalign{\vspace{5pt}}\hline\hline\noalign{\vspace{5pt}}
\end{tabular}
\end{table}

\section{Discussion and Conclusion}
\label{sect:discussion}
The baryonic mass ($M_{bar}$) of the dark galaxy FAST J0139+4328 can be calculated using $M_{bar}=M_{\star}+M_{\rm gas}$, where $M_{\rm gas}$ is the total gas mass.  We do not consider the contribution of stellar mass ($M_{\star}$) to baryonic mass for the galaxy FAST J0139+4328 because it does not have an optical counterpart. Assuming the same helium-to-\HI ratio as that derived from the Big Bang nucleosynthesis, a factor of 1.33 is included to account for the contribution of helium. The total gas mass is determined with $M_{\rm gas} = 1.33\times M_{\rm HI}$. We calculated the $M_{bar}$ of FAST J0139+4328 to be (1.1$\pm$0.3)$\times10^{8} \Msun$, indicating that  FAST J0139+4328 could be  a gas-rich dwarf galaxy. While the dynamic mass of galaxies can be estimated with $M_\mathrm{dyn} = (V^{2}_{\rm rot}+3\sigma_{v}^{2})R_{\rm eff}/G$ \citep{Hoffman1996}, where $V_{\rm rot}$ is rotation velocity, $\sigma_{v}$ is velocity dispersion, $R_{\rm eff}$ is effective radius, and $G$ is gravitational constant. To correct for beam smearing effects, the effective radius can be calculated with $R_{\rm eff} = \sqrt{S_{\rm F}^{2}-B_{\rm F}^{2}}$/2, where  $B_{\rm F}$  is the gridded beam size (3.1$^{\prime}$), and $S_{\rm F}$ is the uncorrected \HI sizes of galaxies.  We adopt the outmost size at 3$\sigma$ level as $S_{\rm F}$ for FAST J0139+4328 from Fig. \textcolor{blue}{1c}. The obtained $R_{\rm eff}$ is 23.9$\pm$2.4 kpc, which appears to be close to the scale of our Galaxy.

To estimate the $V_{\rm rot}$ of FAST J0139+4328, we use the Tilted Ring Fitting Code (TiRiFiC) software package to fit the observed \HI cube data. The TiRiFiC  is a freely available 3D tilted-ring fitting code  \citep{Jozsa2007}, which is a very successful approach to describe the kinematics and morphology of rotating disks. The bootstrap method is used to estimate the final values and errors of model parameters, which are listed in Table \ref{tab:prop}. We also derived the model cube data. In Fig. \textcolor{blue}{1c}, we show the PV diagram from the TiRiFiC model data for FAST J0139+4328 in the blue dashed contours.  Through comparison, it shows that our model can construct the dynamic structure of FAST J0139+4328, except for the weak flat components.  Using the fitted $V_{\rm rot}$ and $\sigma_{v}$, we calculated that $M_{\rm dyn}$ is (5.1$\pm$2.8)$\times10^{9} \Msun$, which is 47$\pm$27 times its baryonic masses, implying that FAST J0139+4328 is dominated by dark matter within the error range. Here, due to the low resolution of the FAST, we can only roughly estimate the content of dark matter. In the near future, we will apply for the higher-angle resolution observation for FAST J0139+4328.

\begin{figure}
\centering
\includegraphics[width=0.40\textwidth]{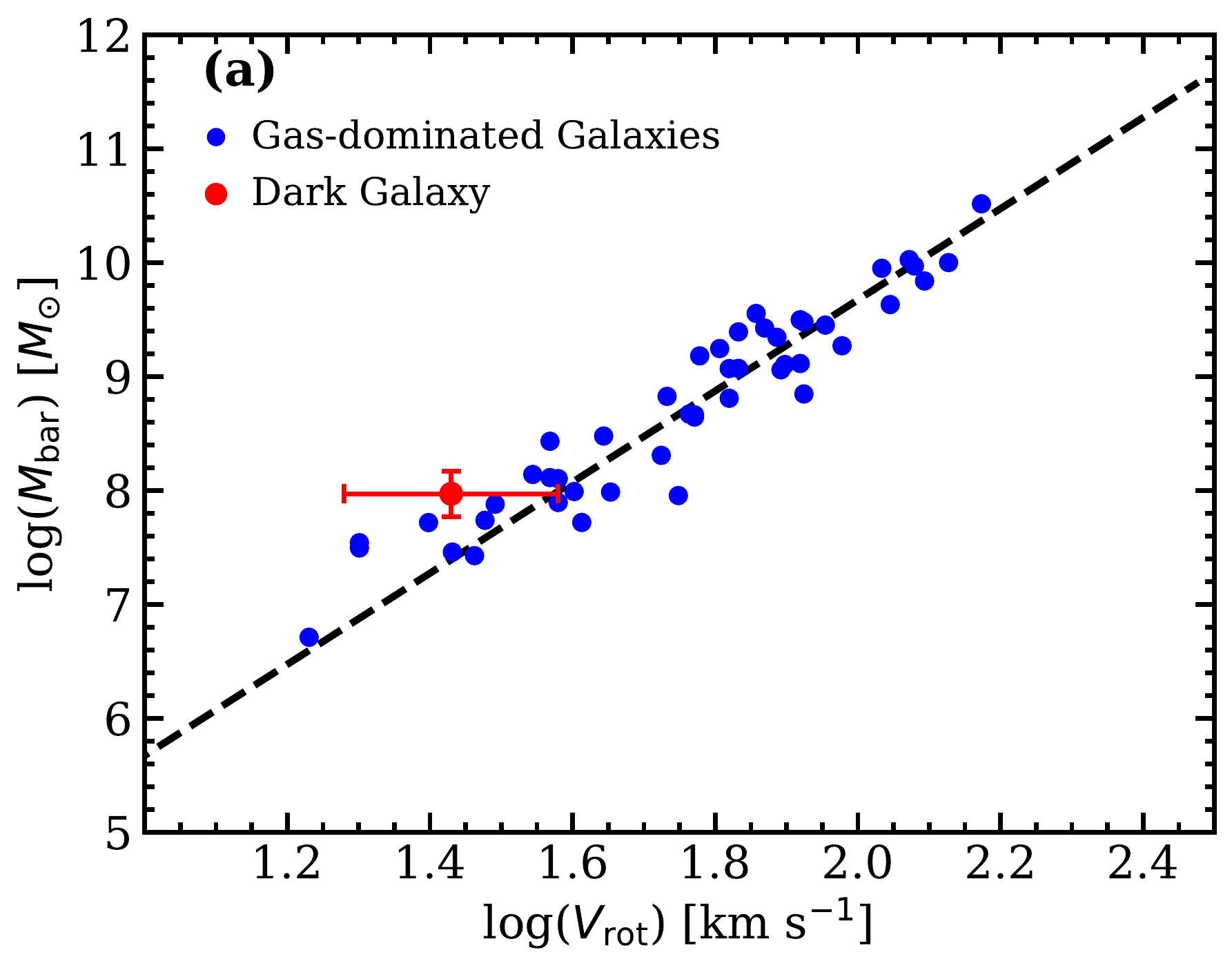}
\includegraphics[width=0.42\textwidth]{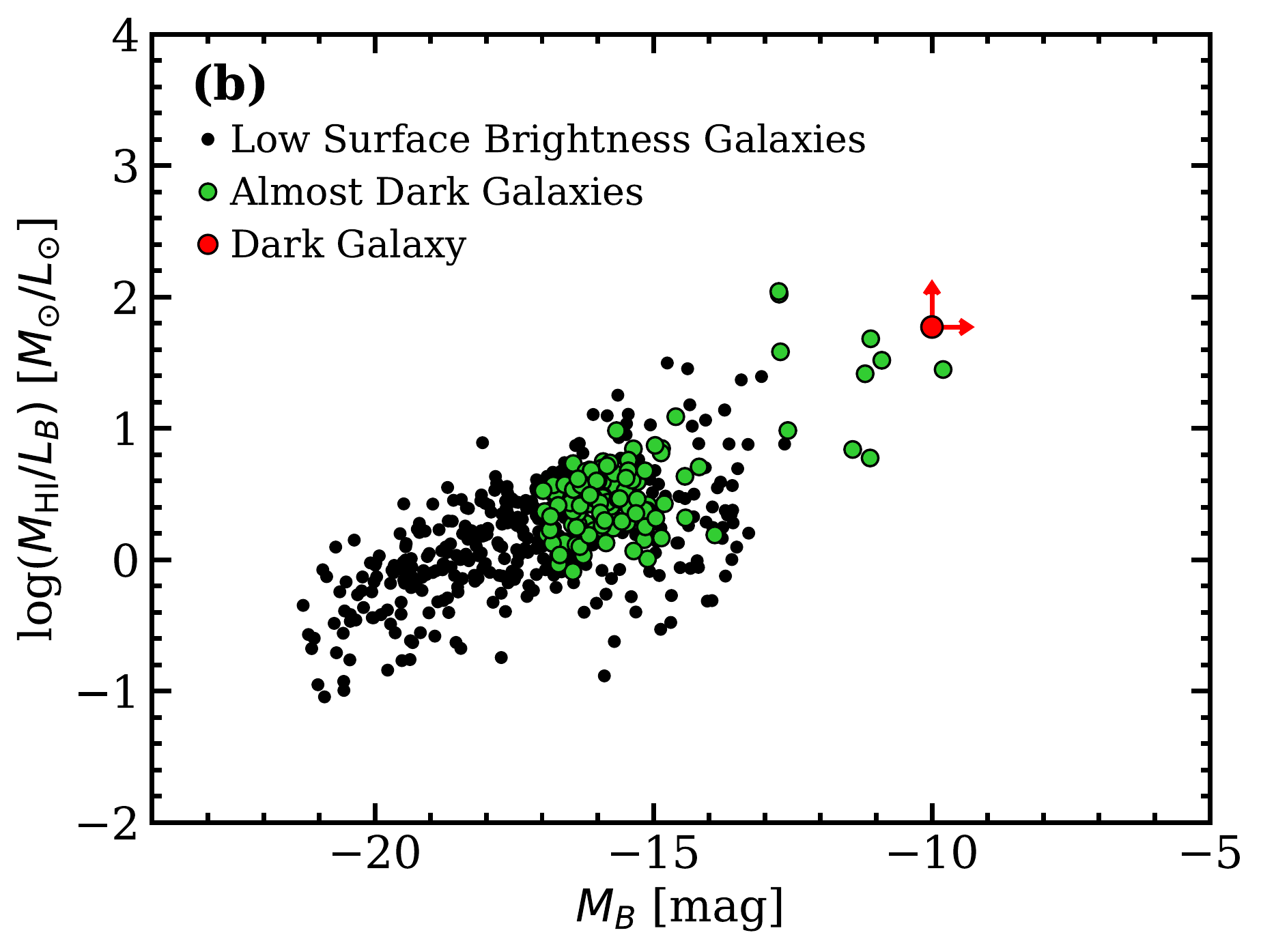}
\caption{(a), Baryonic Tully-Fisher relation (BTFR). The black dashed line represents a best-fit relation $M_{bar}=AV_{\rm rot}^{4}$ with $A=47\pm 6~\msol \rm km^{-4}s^{4}$, and  gas-dominated galaxies are marked by black points \citep{McGaugh2012}. The red point represents the discovered  dark galaxy FAST J0139+4328. (b), relation between the \HI mass-to-light ratio and absolute B-band magnitude. The low surface brightness galaxies are obtained from \citet{Honey2018}, while  almost dark galaxies from \citet{Cannon2015}, \citet{Janowiecki2015}, \citet{Brunker2019} and \citet{Leisman2017,Leisman2021}.}
\label{fig:Tully–Fisher}
\end{figure}

The Baryonic Tully-Fisher relation (BTF) is one of the strongest empirical correlations for disk galaxies. FAST J0139+4328  has no optical counterpart, suggesting that it is gas-dominated. We used the BTF to explore the properties of this special galaxy.  There is a best-fit relation $M_{bar}=AV_{\rm rot}^{4}$ with $A=47\pm 6~\msol \rm km^{-4}s^{4}$ for gas-dominated disk galaxies \citep{McGaugh2012}. Figure \textcolor{blue}{3a}  shows the location of FAST J0139+4328 on the BTF relation.  When compared to other gas-dominated galaxies, FAST J0139+4328 nearly follows the best-fit relation, implying that this galaxy is also a normal disk galaxy. Nevertheless, the disk galaxy FAST J0139+4328 has no optical counterpart, so the internal extinction from dust should be relatively minimal.  We can rule out the possibility that  the internal extinction  is the cause of the absence of an optical counterpart. Moreover, FAST J0139+4328 ($l=132.097^{\circ}$, $b=-18.548^{\circ}$)  is away from the Galactic disk and nearly located in the direction of the anti-Galactic center.   Using a new three-dimensional map of dust reddening \citep{Green2019}, we determined that the dust reddening of $E(B-V)$ is 0.08 in region of FAST J0139+4328. Because of the smaller reddening,  the lack of an optical counterpart for FAST J0139+4328 is also not due to the interstellar extinction.

One reason for the lack of an optical counterpart is the detection limitation  of the Pan-STARRS1 data we used. There is a $B$-band Tully-Fisher relation for spiral galaxies \citep{Tully1977,Zwaan1995}, which is the most commonly used method for determining galaxy distances of redshifts. For FAST J0139+4328, we use the relation $M_{\rm BT}$=$-7.27\times$log$_{10}\it W^{c}_{\rm 20}$+1.94 for spiral galaxies \citep{Tully2000}. The line width is corrected for the inclination angle ($\it i$) as $W^{c}_{20}=W_{\rm 20}$/sin($i$).  Accordint to the $B$-band TF relation, FAST J0139+4328 should have a $M_{\rm BT}$ of -13.0$\pm$0.4 mag if it is a disk galaxy. Nevertheless, FAST J0139+4328 has no optical counterpart. Here we can give an upper limit apparent magnitude for FAST J0139+4328. Using the software SEXtractor \citep{Bertin1996} and a row-by-row and column-by-column method \citep{Wu2002,Du2015}, we created a new sky-background subtraction.

For the optically dark FAST J0139+4328, it needs to set a reasonable aperture diameter. For a very faint optical galaxy, the flux from the sky background dominates the measurement \citep{Leisman2021}. They suggested that it should use a much smaller aperture to measure the flux of a faint galaxy, in order to try to limit the contribution from the sky background. The expected $D_{25}$ of FAST J0139+4328 is 21$^{\prime\prime}$. We adopted $D_{25}$/2 ($\sim$1300 pixels) as an aperture diameter for FAST J0139+4328. From the newly subtracted-background g-band and r-band images, we measured that the 3$\sigma$ detection thresholds are  21.7  mag ($m_{\rm r}$) and 22.0 mag ($m_{\rm g}$) in r-band and g-band  for FAST 0134+5737, which are corrected for Galactic extinction by a new three-dimensional map of dust reddening. Using the transformations for the measured magnitudes from the g-band and r-band to the B-band, the upper limit apparent B-band magnitude can be determined by the equation $m_{\rm B} = m_{\rm g} +0.47(m_{\rm g}-m_{\rm r})+0.17$ \citep{Smith2002}. Then we derived that the upper limit absolute B-band magnitude ($M_{\rm B}$) is -10.0 mag, which is several magnitudes larger than that predicted by the B-band TF relation. It shows that if there is an optical counterpart in the dark galaxy FAST J0139+4328, the Pan-STARRS1 survey images should  be able to reveal the optical emission of the counterpart.

To compare with other faint galaxies, we calculated the \HI mass-to-light ratio ($M_{\rm HI}/L_{\rm B}$). The B-band luminosity can be determined by $L_{\rm B}$=$D^{2}10^{10-0.4(m_{\rm B}-M_{\rm B0})}$, where $M_{B0}$ is the absolute solar  B-band  magnitude, which is adopted as 5.44 mag \citep{Willmer2018}. Using the $M_{\hi}$ of (8.3$\pm$1.7)$\times10^{7} \Msun$, we obtained that $M_{\rm HI}/L_{B}$ is $\geq$59 $M_{\odot}/L_{\odot}$ for FAST J0139+4328, which is much higher than those of 0.15 to 4.2 found in more normal galaxies \citep{Brunker2019}.  Figure~\textcolor{blue}{3b} shows  a relationship between $M_{\rm HI}/L_{\rm B}$ and $M_{\rm B}$. We see that FAST J0139+4328 is almost the darkest gas-rich galaxy. Furthermore, using the g-r color value and B-band luminosity \citep{Zhang2017}, we estimated that the upper limit stellar mass of FAST J0139+4328 is 6.9$\times10^{5} \Msun$. This is the first time that a gas-rich isolated dark galaxy has been detected in the nearby universe. In addition, a galaxy is assumed to form from gas, which cools and turns into stars at the center of a halo. FAST J0139+4328 has a rotating disk of gas and is dominated by dark matter, but is starless, implying that this dark galaxy may be in the earliest stage of the galaxy formation. According to a prior model, however, most of the gas in dark galaxies with baryonic masses greater than  $10^{9}$ \msol will inevitably become Toomre unstable and give rise to star formation in the absence of an internal radiation field \citep{Taylor2005}. We calculated the baryonic mass of FAST J0139+4328 to be (1.1$\pm$0.3)$\times10^{8} \Msun$, indicating that this dark galaxy can endure for a considerable amount of time before being discovered. As a result, we can now detect this dark galaxy. Future blind \HI surveys with high sensitivity and high velocity resolution are expected to  contribute considerably to our understanding for the absence of ultra-faint galaxies.

\acknowledgments 
We thank the referee for insightful comments that improved the clarity of this manuscript. We acknowledge the supports of the National Key R$\&$D Program of China No. 2022YFA1602901. This work is also supported by the Youth Innovation Promotion Association of CAS, the National Natural Science Foundation of China (Grant No. 11933011), the Central Government Funds for
Local Scientific and Technological Development (No. XZ202201YD0020C), and supported by the Open Project Program of the Key Laboratory of FAST, NAOC, Chinese Academy of Sciences.


\begin{thebibliography}{}

\bibitem[{{Bertin} \& {Arnouts}(1996)}]{Bertin1996}
{Bertin}, E., \&  {Arnouts}, S., 1996, AAS,  117, 393

\bibitem[{{Bullock} \& {Boylan-Kolchin}(2017)}]{Bullock2017}
{Bullock}, J. S., \&  {Boylan-Kolchin}, M., 2017, ARA\&A,  55, 343

\bibitem[{{e.g.} {Ball} {et~al.}(2018)}]{Ball2018}
{Ball}, C., {Cannon}, J. M., {Leisman}, L., et al. 2018, \aj, 155, 65.

\bibitem[{{Behroozi}  {et~al.}(2013)}]{Behroozi2013}
{Behroozi}, P. S., {Marchesini}, D., {Wechsler}, R. H., et al. 2013, MNRAS, 777, L10.

\bibitem[{{Broeils} \& {Sancisi}(1997)}]{Broeils1997}
{Broeils}, A. H., \&  {Rhee}, M. H. 1997, \aap,  324, 877

\bibitem[{{Brunker}  {et~al.}(2019)}]{Brunker2019}
{Brunker}, S. W., {McQuinn}, K. B. W., {Salzer}, J. J., et al. 2019, \aj, 157, 76.

\bibitem[{{Cannon}  {et~al.}(2015)}]{Cannon2015}
{Cannon}, J. M., {Martinkus}, C. P., {Leisman}, L., et al. 2015, \aj, 149, 72.

\bibitem[{{Chambers}  {et~al.}(2016)}]{Chambers2016}
{Chambers}, K. C., {Magnier}, E. A., {Metcalfe}, N., et al. 2016, arXiv:1612.05560.

\bibitem[{{Davies}  {et~al.}(2006)}]{Davies2006}
{Davies}, J. I., {Disney}, M. J., {Minchin}, R. F., et al. 2006, MNRAS, 368, 1479.

\bibitem[{{Doyle}  {et~al.}(2005)}]{Doyle2005}
{Doyle}, M. T., {Drinkwater}, M. J., {Rohde}, D. J., et al. 2005, MNRAS, 361, 34.

\bibitem[{{Du}  {et~al.}(2015)}]{Du2015}
{Du}, W., {Wu}, H., {Lam}, M. I., et al. 2015, \aj, 149, 199.

\bibitem[{{Duc} \& {Bournaud}(2008)}]{Duc2008}
{Duc}, P. A., \&  {Bournaud}, F,. 2008, \aj,  673, 787

\bibitem[{{Gaia}  {et~al.} (2016)}]{Gaia2016}
{Gaia Collaboration}., {Prusti}, T., {de Bruijne}, J. H. J., et al. 2016, \aap, 595, A1.

\bibitem[{{Green}  {et~al.} (2019)}]{Green2019}
{Green}, G. M., {Schlafly}, E., {Zucker}, C., et al. 2019, \apj, 887, 93.

\bibitem[{{Glowacki}  {et~al.} (2022)}]{Glowacki2022}
{Glowacki}, M., {Collier}, J. D., {Kazemi-Moridanl}, A., et al. 2022, \apj, 931, L7.

\bibitem[{{Haynes}  {et~al.}(2011)}]{Haynes2011}
{Haynes}, M. P., {Giovanelli}, R., {Martin}, A. M., et al. 2011, \aj, 142, 170.


\bibitem[{{Hess}  {et~al.}(2021)}]{Hess2021}
{Hess}, K. M., {Roberts}, H., {D{\'e}nes}, H., et al. 2021, \aap, 647, 193.

\bibitem[{{Hoffman}  {et~al.}(1996)}]{Hoffman1996}
{Hoffman}, G. L., {Salpeter}, E. E., {Farhat}, B., et al. 1996, \apjs, 105, 269.

\bibitem[{{Honey}  {et~al.}(2018)}]{Honey2018}
{Honey}, M., {van Driel}, W., {Das}, M., \& {Martin}, J. M., 2018, MNRAS, 476, 4488.

\bibitem[{{Janowiecki}  {et~al.}(2015)}]{Janowiecki2015}
{Janowiecki}, S., {Leisman}, L., {J{\'o}zsa}, G.,  et a., 2015, \apj, 801, 96.

\bibitem[{{FAST,} {Jiang} {et~al.}(2019)}]{Jiang2019}
{Jiang}, P., {Yue}, Y. L., {Gan}, H. Q., et al. 2019, Sci. China-Phys.Mech. Astron. 62, 959502

\bibitem[{{Jiang} {et~al.}(2020)}]{Jiang2020}
{Jiang}, P., {Tang}, N.-Y., {Hou}, L.-G., et al. 2020, Research in Astronomy and Astrophysics, 20, 064

\bibitem[{{J{\'o}zsa}  {et~al.}(2017)}]{Jozsa2007}
{J{\'o}zsa}, G. I. G., {Kenn}, F., {Klein}, U., {Oosterloo}, T. A., 2007, \aap, 468, 731.

\bibitem[{{Kauffmann}  {et~al.}(1993)}]{Kauffmann1993}
{Kauffmann}, G., {White}, S. D. M., \& {Guiderdoni}, B. 1993, MNRAS, 264, 201.

\bibitem[{{Kourkchi}  {et~al.}(2020)}]{Kourkchi2020}
{Kourkchi}, E., {Courtois}, H. M., {Graziani}, R., et al. 2020, \aj, 159, 67.

\bibitem[{{Leisman}  {et~al.}(2017)}]{Leisman2017}
{Leisman}, L., {Haynes}, M. P., {Janowiecki}, S., et al. 2017, \apj, 842, 133.

\bibitem[{{Leisman}  {et~al.}(2021)}]{Leisman2021}
{Leisman}, L., {Rhode}, K. L., {Ball}, C., et al. 2021, \aj, 162, 274.

\bibitem[{{Martin}  {et~al.}(2005)}]{Martin2005}
{Martin}, D. C., {Fanson}, J., {Schiminovich}, D., et al. 2005, \apj, 619, L1.

\bibitem[{{McGaugh} (2012)}]{McGaugh2012}
{McGaugh}, S. S., 2012, \aj, 143, 40.

\bibitem[{{Moore}  {et~al.}(1999)}]{Moore1999}
{Moore}, B., {Ghigna}, S., {Governato}, F., et al. 1999, \apj, 524, L19.

\bibitem[{{Roman}  {et~al.}(2021)}]{Roman2021}
{Rom{\'a}n}, J., {Jones}, M. G., {Montes}, M., et al. 2021, \aap, 649, L14.

\bibitem[{{Sawala}  {et~al.}(2016)}]{Sawala2016}
{Sawala}, T., {Frenk}, C. S., {Fattahi}, A., et al., 2016, MNRAS, 457, 1931.

\bibitem[{{Sales}  {et~al.}(2022)}]{Sales2022}
{Sales}, L. V., {Wetzel}, A., \& {Fattahi}, A. 2022, Nature Astronomy, 6, 897.

\bibitem[{{Skrutskie}  {et~al.}(2006)}]{Skrutskie2006}
{Skrutskie}, M. F., {Cutri}, R. M., {Stiening}, R., et al., 2006, \aj, 131, 1163.

\bibitem[{{Smith}  {et~al.}(2002)}]{Smith2002}
{Smith}, J. A., {Tucker}, D. L., {Kent}, S.,  et al. 2002, \aj, 123, 2121.

\bibitem[{{Solanes}  {et~al.}(1996)}]{Solanes1996}
{Solanes}, J. M., {Giovanelli}, R., \& { Haynes}, M. P. 1996, \apj, 461, 609.

\bibitem[{{Taylor} \& {Sancisi}(2005)}]{Taylor2005}
{Taylor}, E. N., \&  {Webster}, R. L. 2005, \apj,  634, 1067

\bibitem[{{Taylor}  {et~al.}(2016)}]{Taylor2016}
{Taylor}, R., {Davies}, J. I., {J{\'a}chym}, P., et al. 2016, MNRAS, 461, 3001.

\bibitem[{{Taylor}  {et~al.}(2017)}]{Taylor2017}
{Taylor}, R. {Davies}, J. I., {J{\'a}chym}, P., et al. 2017, MNRAS, 467, 3648.

\bibitem[{{Toribio}  {et~al.}(2011)}]{Toribio2011}
{Toribio}, M. C., {Solanes}, J. M., {Giovanelli}, R., et al. 2011, \apj, 732, 93.

\bibitem[{{Tollerud} {et~al.}(2008)}]{Tollerud2008}
{Tollerud}, E. J., {Bullock}, J. S., {Strigari}, L. E., \& {Willman}, B. 2008, \apj, 688, 277

\bibitem[{{Tully} \& {Sancisi}(1977)}]{Tully1977}
{Tully}, R. B., \&  {Fisher}, J. R. 1977, \aap,  54, 661

\bibitem[{{Tully} \& {Pierce}(2000)}]{Tully2000}
{Tully}, R. B., \&  {Pierce}, M. J. 2000, \apj,  533, 744

\bibitem[{{Turner} \& {MacFadyen}(1997)}]{Turner1997}
{Turner}, N. J. J., \&  {MacFadyen}, A. A., 1997, MNRAS,  285, 125

\bibitem[{{Willmer}(2018)}]{Willmer2018}
{Willmer}, C. N. A., 2018, \apjs, 236, 47

\bibitem[{{Westmeier} {et~al.}(2014)}]{Westmeier2014}
{Westmeier}, T., {Jurek}, R., {Obreschkow}, D. et al. 2014, MNRAS, 438, 1176

\bibitem[{{Wu} {et~al.}(2002)}]{Wu2002}
{Wu}, H., {Burstein}, D., {Deng}, Z. G., et al. 2002, \aj, 123, 1365

\bibitem[{{Xu} {et~al.}(2021)}]{Xu2021}
{Xu}, J. L., {Zhang}, C. P., {Yu}, N. et al. 2021, \apj, 922, 53

\bibitem[{{Zwaan} {et~al.}(1995)}]{Zwaan1995}
{Zwaan}, M. A., {van der Hulst}, J. M., {de Blok}, W. J. G., \& {McGaugh}, S. S., 1995, MNRAS, 273, L35

\bibitem[{{Zhang} {et~al.}(2021)}]{Zhang2021}
{Zhang}, C. P., {Xu}, J. L., {Li}, G. et al. 2021, RAA, 21, 209

\bibitem[{{Zhang} {et~al.}(2022)}]{Zhang2022}
{Zhang}, C. P., {Xu}, J. L., {Wang}, J. et al. 2022, RAA, 22, 10

\bibitem[{{Zhang}  {et~al.}(2017)}]{Zhang2017}
{Zhang}, H. X., {Puzia}, T., \& {Weisz}, D. R., 2017, \apjs, 233, 12.

\bibitem[{{Zhang} {et~al.}(2014)}]{Zhang2014}
{Zhang}, J. S., {Wang}, J. Z., {Di}, G. X., et al. 2014, \aap, 570, A110

\end{thebibliography}
\end{document}